\def\selectedoptions{final}
\ifx\empty\selectedoptions
  \def\selectedoptions{final}
\fi
\documentclass[
   \selectedoptions
  ]
  {aipproc}

\def\selectedlayoutstyle {8x11double}
\layoutstyle\selectedlayoutstyle

\SetInternalRegister\hbadness{8000} 

\newcommand\doingARLO[2][]{%
  \ifx\mmref\undefined #1\else #2\fi
}
  
\begin{document}

\def\apj{{\it Ap.\ J.,}}
\def\apjl{{\it Ap.\ J. Letters,}}
\def\aa{{\it Astron.\ Astrophys.,}}

\title 
      [In-flight verification of the FREGATE spectral response]
      {In-flight verification of the FREGATE spectral response}


\author{J-F. Olive}{
  address={Centre d'Etude Spatiale des Rayonnements, CNRS/UPS, 31028 Toulouse Cedex 04, France},
  email={olive@cesr.fr},
}

\author{J-P. Dezalay}{
  address={Centre d'Etude Spatiale des Rayonnements, CNRS/UPS, 31028 Toulouse Cedex 04, France},
}
\author{J-L. Atteia}{
  address={Centre d'Etude Spatiale des Rayonnements, CNRS/UPS, 31028 Toulouse Cedex 04, France},
}
\author{C. Barraud}{
  address={Centre d'Etude Spatiale des Rayonnements, CNRS/UPS, 31028 Toulouse Cedex 04, France},
}

\author{N. Butler}{
  address={Massachusetts Institute of Technology, Center for Space Research, Cambridge, MA, US},
}
\author{G. B. Crew}{
  address={Massachusetts Institute of Technology, Center for Space Research, Cambridge, MA, US},
}
\author{J. Doty}{
  address={Massachusetts Institute of Technology, Center for Space Research, Cambridge, MA, US},
}
\author{G. Ricker}{
  address={Massachusetts Institute of Technology, Center for Space Research, Cambridge, MA, US},
}
\author{R. Vanderspek}{
  address={Massachusetts Institute of Technology, Center for Space Research, Cambridge, MA, US},
}

% \copyrightholder{Acoustical Scociety of America}
\copyrightyear  {2001}

\begin{abstract}

We present the first results of the in-flight validation of the
spectral response of the FREGATE X/$\gamma$ detectors on--board the
HETE--2 satellite. This validation uses the Crab pulsar and nebula as
reference spectra.

\end{abstract}

\date{\today}

\maketitle

\section{Introduction}

\ifthenelse{\equal\selectedlayoutstyle{6x9}}{\par\bfseries 
  Note: The entire paper will be reduced 15\% in the printing
  process. Please make sure all figures as well as the text within the
  figures are large enough in the manuscript to be readable in the
  finished book.\par\bfseries 
  Note: The entire paper will be reduced 15\% in the printing
  process. Please make sure all figures as well as the text within the
  figures are large enough in the manuscript to be readable in the
  finished book.\par\bfseries 
  Note: The entire paper will be reduced 15\% in the printing
  process. Please make sure all figures as well as the text within the
  figures are large enough in the manuscript to be readable in the
  finished book.\normalfont}{}

The four identical FREGATE X/$\gamma$--ray detectors onboard HETE-2,
are sensitive to photons between 6 and 400 keV (see
\cite{atteia-wh} for a full description of the experiment and
operating modes). In this short paper, we describe briefly the Monte
Carlo simulations of the detectors, the ground calibrations, and the
validation of the in-flight performances using the Crab pulsar and
nebula as standards. We compare the spectral parameters obtained with
FREGATE with those reported by other instruments in similar energy
ranges.

\section{Simulations and calibrations}

\begin{figure} [t!]
  \resizebox{0.5\columnwidth}{!}  {\includegraphics{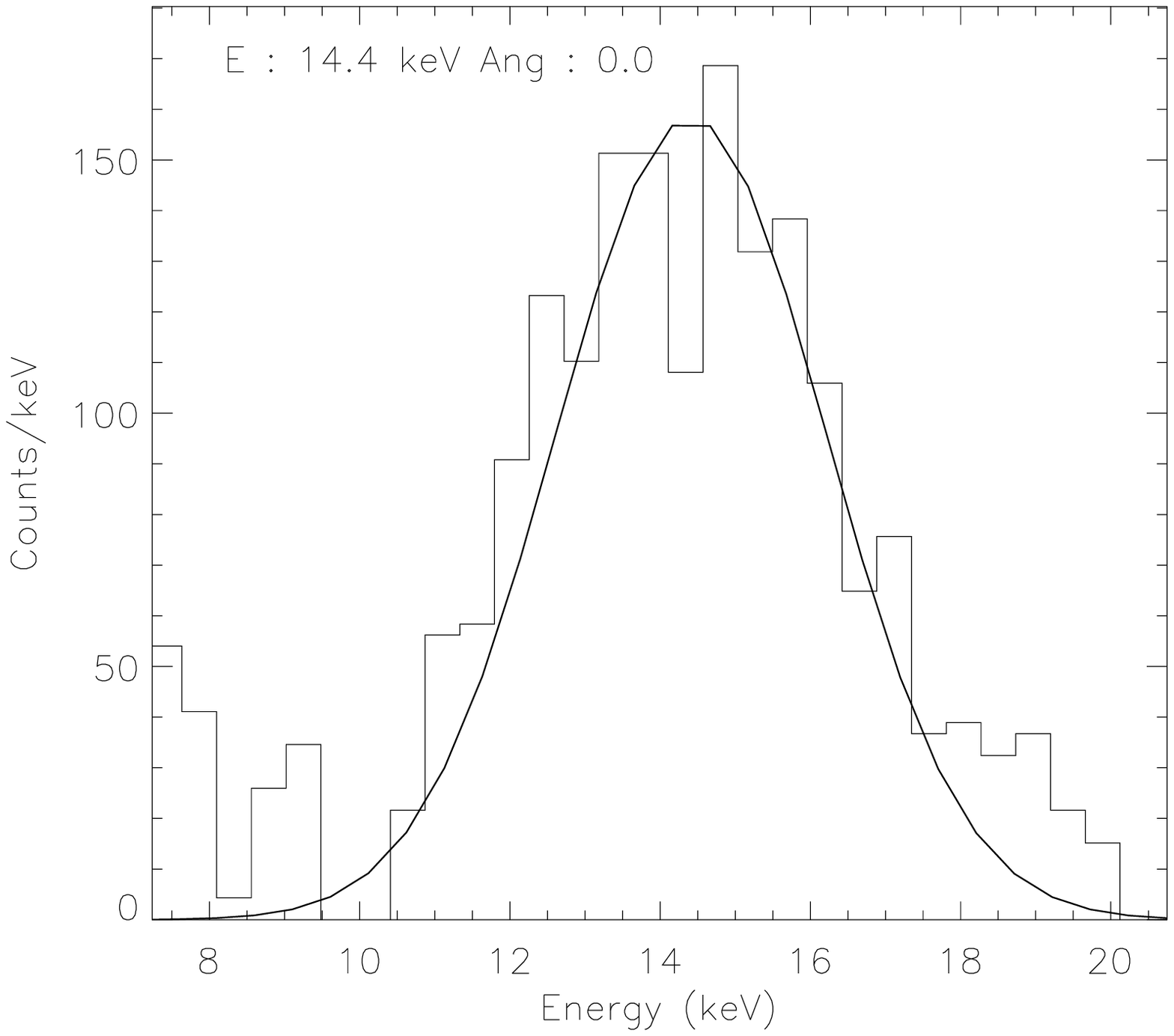}}
  \resizebox{0.5\columnwidth}{!}  {\includegraphics{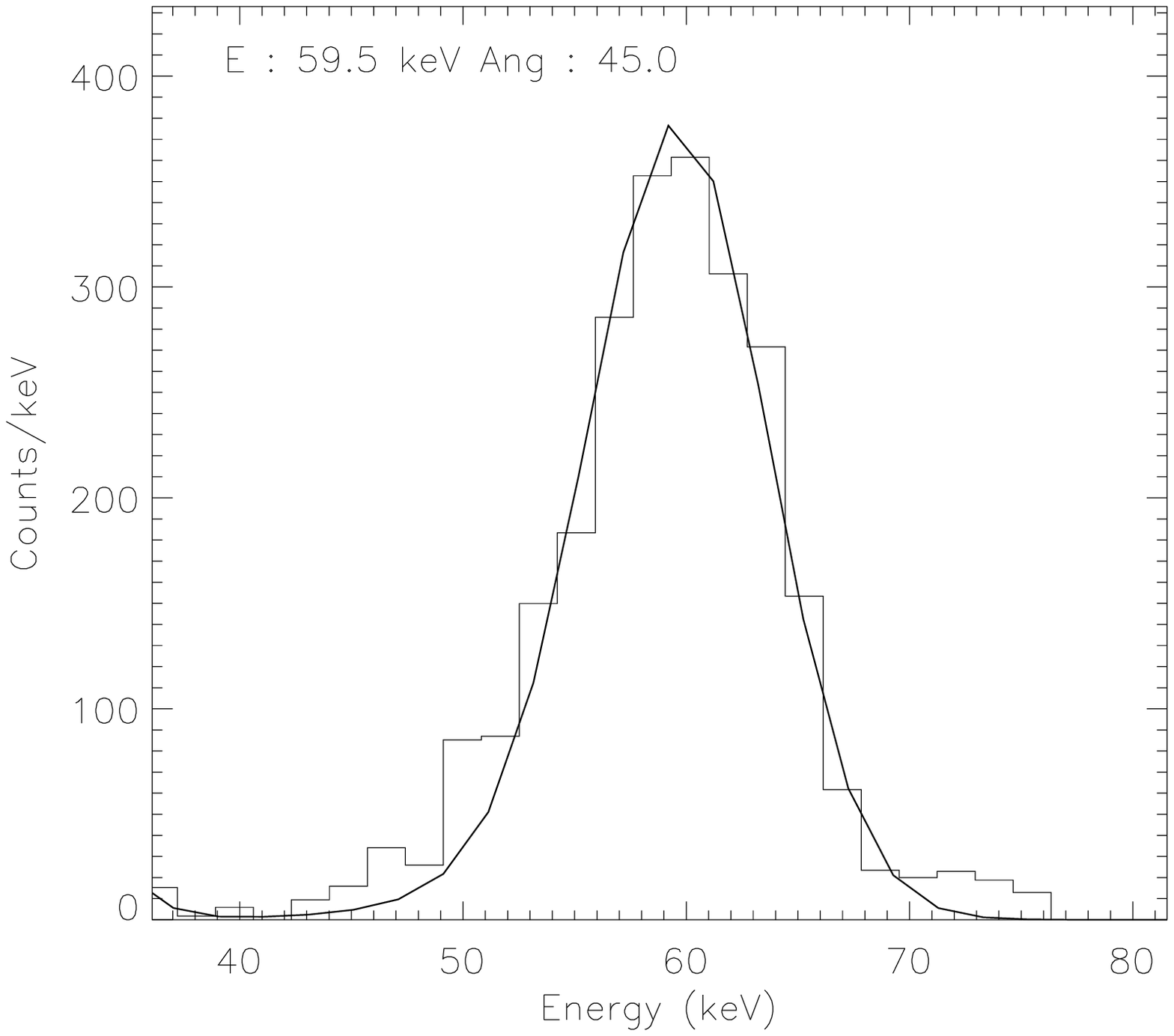}}
  \resizebox{0.5\columnwidth}{!}  {\includegraphics{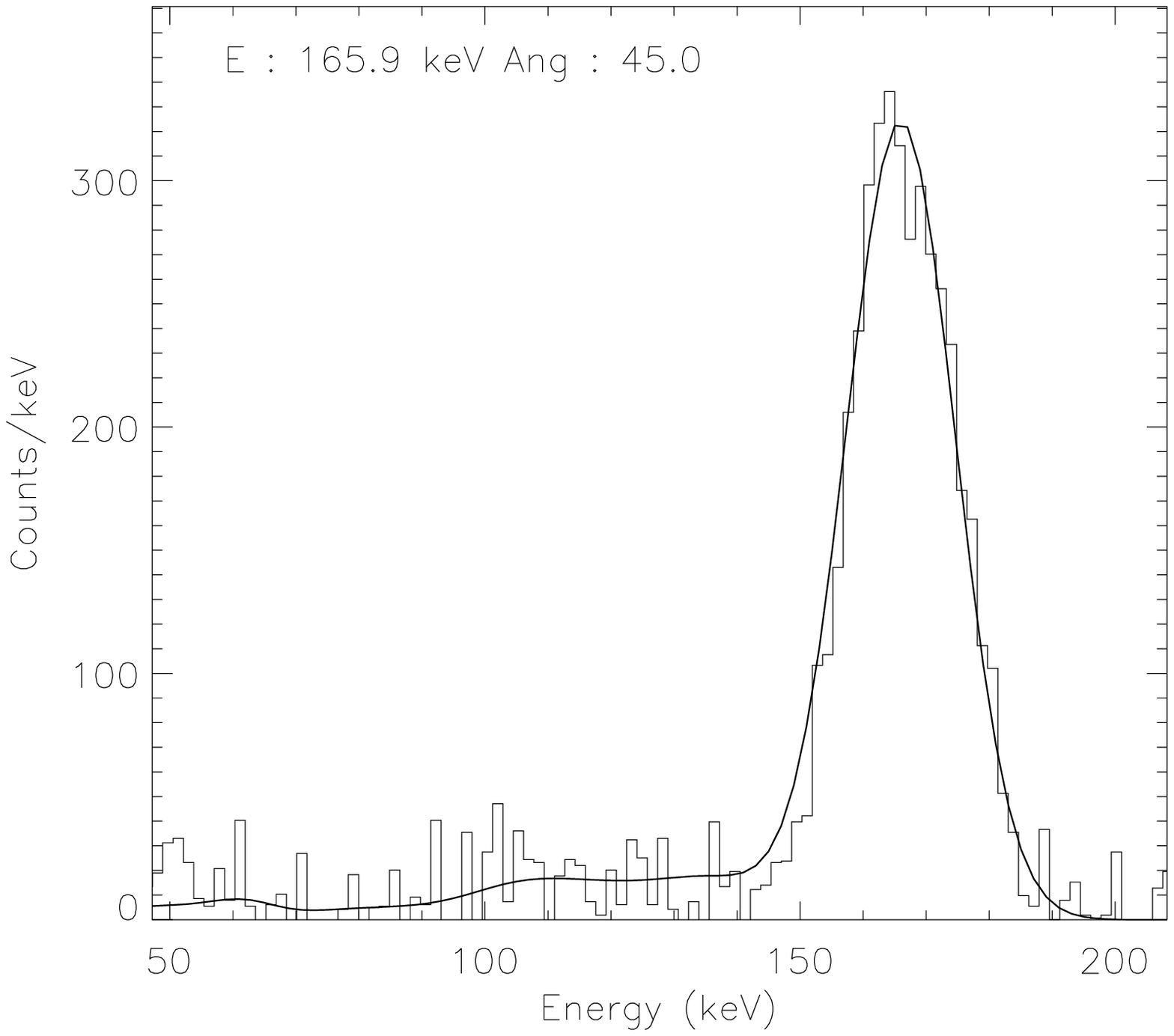}}
  \resizebox{0.5\columnwidth}{!}  {\includegraphics{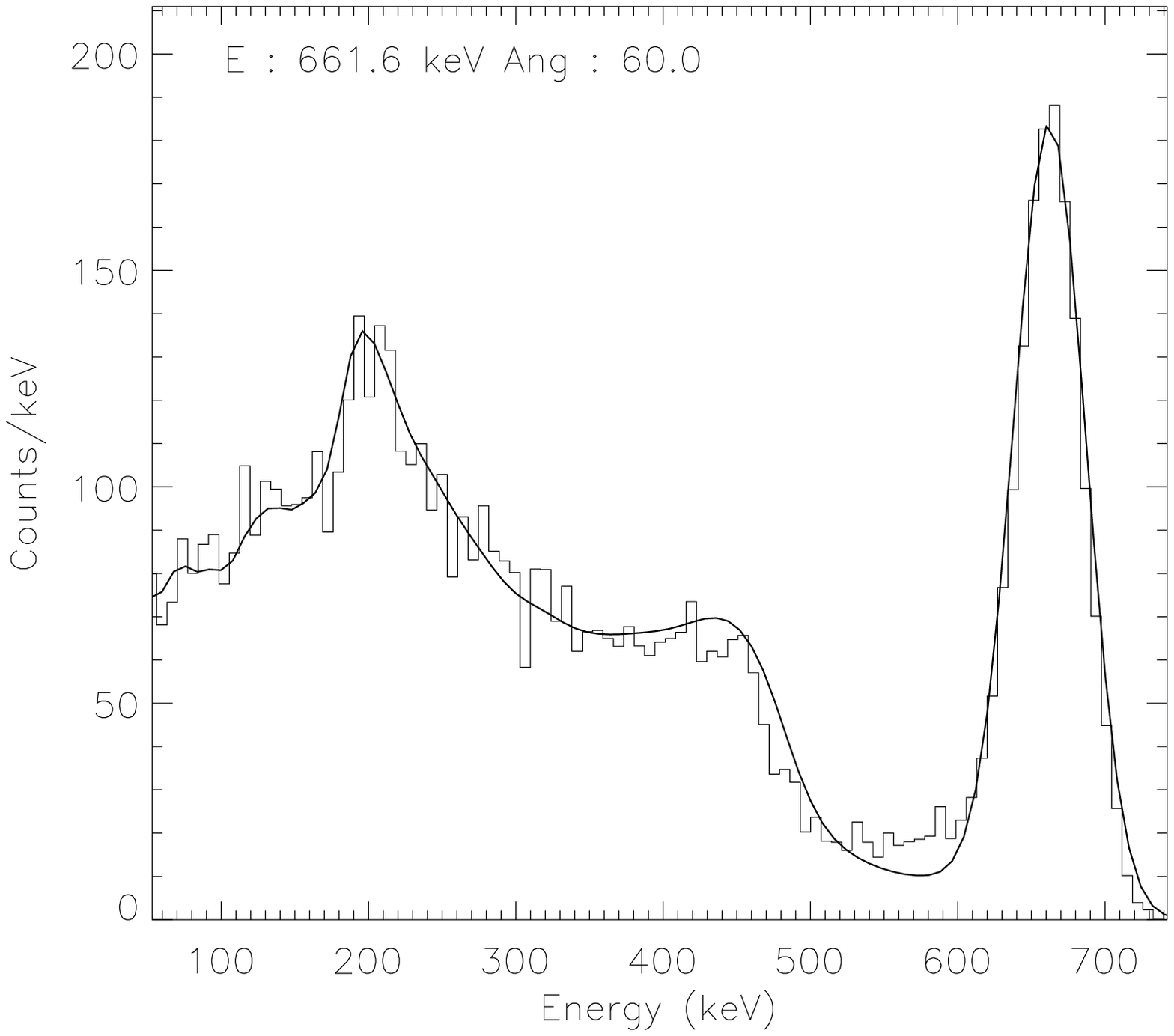}}

  \caption{Comparison of experimental spectra of several radioactive
  sources from ground calibrations (histograms) and the corresponding
  simulated spectra (solid lines). From left to right : $^{60}$Co (14
  keV, 0$^{\circ}$), $^{241}$Am (59 keV, 45$^{\circ}$), $^{139}$Ce
  (166 keV, 45$^{\circ}$) and $^{137}$Cs (662 keV, 60$^{\circ}$). The
  simulations are normalized using the source activity for the given
  energy, the duration of the calibration run, and the solid angle of
  the detector seen by the source. In other words, it is an absolute
  comparison.}

  \label{calib_plot}
\end{figure}

In parallel with the construction of the detectors, the energy
response matrix of FREGATE was calculated from extensive Monte Carlo
simulations of the detector using CERN's GEANT package. More than 50
regions of different materials were included in the simulation with
special care for the regions in the detector field-of-view (e.g. the
graded shield and the beryllium window). The parameters (position,
angles, energy, etc.)  of the incoming photons are generated within a
separate code including several options (point-like radioactive source
at a finite distance with several photons and branching ratios,
parallel flux for celestial sources, etc.). Once the photon
characteristics are generated, the tracking code is the same in all
cases. The Monte-Carlo simulations were checked before launch with the
help of ground calibrations using a large set of radioactive sources
(9 sources with photons energies spanning from 8 to 1300 keV) and
angles of incidence from on-axis to 60$^{\circ}$ in 5$^{\circ}$
steps. We have first determined realistic gains and resolutions for
our detectors. Then, the simulated spectra (in a point-like
configuration) have been folded with the detector response functions
and normalized knowing the source activity for the given energy, the
duration of the calibration run and the solid angle presented by the
detector to the photon flux. Finally, we compared the simulated
spectra with those from the calibrations. For all angles and energies,
we found that the relative differences in the full energy peak and in
the diffusion continuum were less than 10 \%, which is on the order of
the uncertainties of the source activity (see Figure \ref{calib_plot}
for few examples of comparison). Since we concentrate in this paper on
the validation of the in-flight performances, we will not discuss
these comparisons any further.

\section{In-flight calibrations}

Since launch, the gains of each detector have been continuously
monitored using on-board calibration $^{133}$Ba sources (two lines at
81 and 356 keV). The positions of these lines show a weekly orbital
variation of a few percent, and a long-term drift on a $\sim$ 100 day
scale periodically corrected by adjusting the high voltage
commands. No significant variation of the relative line width has been
found. These in-flight calibrations are used to find the correct
channel-to-energy relations during the HETE mission. FREGATE generates
two types of scientific data with good spectral resolution :
permanently the `Spectral Data' ({\bf SD}: four 128 channel energy
spectra every 5 or 10 seconds); and when a trigger occurs : the `Burst
Data' ({\bf BD}: 256k photons tagged in time with a resolution of 6.4
$\mu$s in time and in 256 channels in energy). The process of spectral
deconvolution for both data types has been tested during the in-flight
verification phase, using the Crab pulsar (for the BD) and nebula (for
the SD). The counting rates in the 6--200 keV range obtained from the
four detectors have been rebinned in the same histogram to improve the
statistical significance. We have checked that fitting the four
spectra simultaneously gives the same spectral parameters within the
error bars. As usual for the Crab, we used a single power law model :

\begin{equation}
\frac{dN}{dE} = A_{30}~E_{30}^{-\gamma}~{\rm ph}~{\rm cm}^{-2}~{\rm s}^{-1}~{\rm keV}^{-1}
\end{equation}
where $E_{30}$ is defined as $E_{30}=E/30$ keV

\subsubsection {The Crab pulsar}

\begin{figure} [h]
  \resizebox{1.\columnwidth}{!}
     {\includegraphics[angle=90]{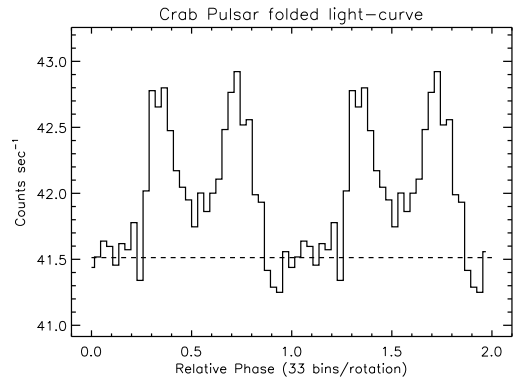}}
  \caption{The Crab pulsar light-curve folded at the best pulsation period. Two periods are shown.}
  \label{lc_plot}
\end{figure}

The data consist of 10 artificial burst triggers recorded on
01/07/2001, when the Crab was about 23$^{\circ}$ off axis. Considering
that the triggers are separated by only a few minutes, the photon
times were not barycenter corrected. We have constructed the folded
light-curves (with a 33 phase bin epoch folding) for periods around
the expected Crab period (\cite{lyne}). For each light-curve, we have
computed the $\chi^2_{red}$ value. For the maximal value of this
statistical parameter ($\chi^2_{red} = 10.1$) we have found the best
pulsation period ($\nu$ =29.832951 Hz). This value is $\sim 10^{-3}$
Hz smaller than the extrapolated value from the ephemeris ($\nu_{0}$
=29.834086 Hz) which is compatible with a Doppler shift due to the
earth's motion in the solar system and the satellite motion in its
orbit. The corresponding phasogram (Figure \ref{lc_plot}) consists of
two peaks of similar intensity, separated by $\sim$ 0.4 in phase and
connected by an interpulse.  Even if the peaks are a little broad (due
to the lack of barycentric corrections), this phasogram looks very
similar to the ones reported at X and $\gamma$ energies. The off-pulse
level and spectrum were obtained for phases outside the pulsed emission
(dashed level, Figure~\ref{lc_plot}).

\begin{table} [t!]
\begin{tabular}{cccccc}
\hline
    \tablehead{1}{c}{b}{Reference}   
  & \tablehead{1}{c}{b}{Energy\\range}
  & \tablehead{1}{c}{b}{Pulsar (phase averaged) \\$A_{30}$}
  & \tablehead{1}{c}{b}{$\gamma$}
  & \tablehead{1}{c}{b}{Nebula (total emission)\\$A_{30}$}
  & \tablehead{1}{c}{b}{$\gamma$} \\

\hline
FREGATE                       & 6-200 keV  & {\bf 0.89 $\pm$ 0.13 } & {\bf 1.87 $\pm$ 0.13} & {\bf 7.23 $\pm$ 0.2}   & {\bf 2.16 $\pm$ 0.03}   \\
\hline
\cite{hubert94}               & 15-130 keV & 1.04                   & 2.06 $\pm$ 0.3        & 7.05                   & 2.18 $\pm$ 0.04   \\
\cite{jung89}                 & 15-180 keV &  --                    &      --               & 7.46                   & 2.06 $\pm$ 0.01   \\
\cite{hasin84}                & 20-200 keV & 1.06                   & 1.92 $\pm$ 0.09       & 7.48                   & 1.94 $\pm$ 0.02   \\
\cite{strick79}               & 20-250 keV & 1.16                   & 1.96 $\pm$ 0.05       & 7.18                   & 2.19 $\pm$ 0.02   \\
\hline
\end{tabular}
\label{tableparamtotal}

 \caption{Summary for Crab pulsar and Crab nebula spectral parameters
 (FREGATE results and \cite{hubert94}). The spectral index is
 represented by $\gamma$. The parameter $A_{30}$ (intensity at 30 keV)
 is in units of $10^{-3}$~ph~cm$^{-2}$~s$^{-1}$~keV$^{-1}$. While
 given by the authors at a different energy, this amplitude has been
 recalculated using the quoted best-fit slope.}

\end{table}

We obtained a good fit with the spectral model described above
($\chi^2_{red} = 1.1$ for 86 dof). The rebinned unfolded spectrum can
be seen in Figure \ref{ufs}. The spectral parameters are reported in
Table \ref{tableparamtotal}. The spectral index (1.87 $\pm$ 0.13) is
fully consistent with those reported in similar energy ranges. The
amplitude ($A_{30}$) is about 10~\% lower than the canonical value
(although roughly consistent within the error bars). This can be
explained because, in absence of timing corrections, the FREGATE light
curve is smoothed and our definition of the total pulsed interval (0.6
in phase) is greater than the one usually used ($\sim$ 0.5-0.55).

\subsubsection {The Crab Nebula}

\begin{figure} [h!]
  \resizebox{.85\columnwidth}{!}
     {\includegraphics[angle=0]{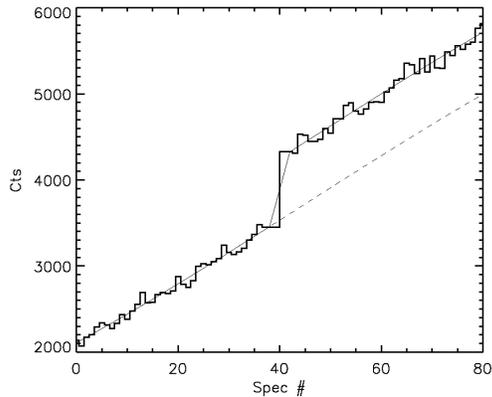}}
  \label{marche}

 \caption{Exemple of light curve obtained by folding the Spectral Data
 of FREGATE with respect to the Crab occultation times. The amplitude
 of the step represents the number of counts due to the Crab nebula.}

\end{figure}

The Crab was periodicaly occulted by the Earth in the FREGATE
field-of-view with an inclination angle less than 10$^{\circ}$ on December,
2000. The occultation dates (Crab rise and Crab set) were
calculated. From the SD, we have extracted sets of 80 successive
spectra centered on the occultation dates. With a screening procedure,
the data polluted with solar flares or high background were
eliminated, leading to 90 `good' occultations. Then we have added all
these 90 occultations (spectrum by spectrum), to get an `averaged' set
of 80 successive spectra centered on the Crab steps (a Crab set is
time-reversed before adding). Next, the light-curve for each of the
128 energy channels has been fitted with a polynomial function of
fourth degree to account for the background orbital variation plus a
centered step to account for the Crab contribution in the channel (see
an exemple in Figure \ref{marche}). Thus, the Crab spectrum is
built channel-by-channel for each of the four FREGATE detectors.

\begin{figure} [h]
  \resizebox{1\columnwidth}{!}
     {\includegraphics[angle=0]{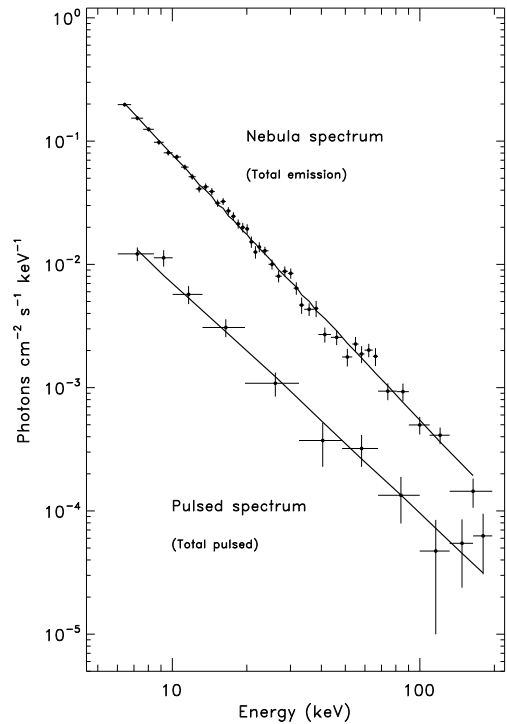}}
   \caption{Unfolded spectrum of the total nebula emission and the phase
 averaged Crab pulsed emission}  
\label{ufs}
\end{figure}

Again, we obtained a good fit with a power law model ($\chi^2_{red} =
1.19$ for 84 dof, $\gamma = 2.16 \pm 0.03$) with spectral parameters
fully consistent with those reported in similar energy ranges (see the
unfolded spectrum in Figure \ref{ufs} and the parameters in Table
\ref{tableparamtotal}).

\section{Conclusion}

For both the Crab pulsar and nebula spectra, the spectral parameters
derived with FREGATE are fully consistent with the canonical
values. This demonstrates our capability to perform a detailed
spectral analysis of sources within the FREGATE field-of-view, even
for low statistics spectra such as the pulsed spectrum presented
here.

\doingARLO[\bibliographystyle{aipproc}]
          {\ifthenelse{\equal{\AIPcitestyleselect}{num}}
             {\bibliographystyle{arlonum}}
             {\bibliographystyle{arlobib}}
          }

\end{document}